\documentclass{elsart1p}
\usepackage{graphicx}
\usepackage{epsfig}
\usepackage{amssymb}
\begin{document}
 \begin{frontmatter}
  \title{Factorization and resummation for color octet production}
\author{ Ahmad Idilbi and Chul Kim}
\address{Physics Department, Duke University, Durham, NC, 27708, USA}
\begin{abstract}
 We discuss the production of heavy colored paricles at the Large Hadron Collider (LHC) through gluon-gluon fusion process.
A factorization theorem is obtained for this process using Soft Collinear Effective Theory. Our factorization theorem does not depend on any assumptions regarding the physics above the mass of the heavy colored particle. In this sense it is universal. The matching coefficient at the heavy particle mass scale depends however on the unkown physics above that scale and thus it is model dependent. 

Due to the large mass of the heavy colored particle, i.e., the hard scale and near the partonic kinematic threshold for production of such particles a resummation of large logarithms needs to be performed. The resummation is justified due to the dominance of the gluon distribution function at small $x$. For phenomenological purposes we utilize the Manohar-Wise model.
\end{abstract}
\begin{keyword}
 Factorization \sep Resummation \sep Effective Theory
\PACS 12.39.St \sep 14.80.-j
\end{keyword}
\end{frontmatter}
A common feature of various extensions of the Standard Model (SM) is the existence of heavy colored scalars as part of their particle spectrum. Such particles have not yet been observed and it is up for the LHC to determine whether they do or do not exist. In this short account we discuss the possible production of heavy octet (under $SU(3)_c$) scalars
at the LHC through gluon-gluon fusion process as was recently proposed by Manohar and Wise \cite{Manohar:2006ga}. The specifics of Manohar-Wise model will not be discussed here. We only mention that in this model the heavy scalar octet will Yukawa couple to the SM fermions much like in the SM Higgs case.

To perform precise phenomenological study of the cross section for production of (single) heavy octet scalars at the LHC one needs to establish a factorization theorem which separates the perturbatively calculable quantities (hard and soft functions) from the non-perturbative ones (parton distribution functions (PDF).) Again this is similar to the more familiar case of the SM Higgs production (via its Yukawa coupling to the top quark) with one major difference which is the color charge carried by the scalar octet. The color exchange between the incoming gluons and the produced scalar octet highly complicates the derivation of such factorization theorem. 

Let $s$ be the total hadronic center of mass (COM) energy squared, $\hat{s}$-the total partonic COM squared, $M_S$-the mass of the produced scalar octet and $z=M_S^2/\hat{s}$. In the limit of partonic threshold, $z \to 1$, our derived factorization theorem reads 
\begin{eqnarray} 
\sigma (pp\to SX) &=& \tau H(M_S,\mu_s) \int \frac{dy_1}{y_1} \frac{dy_2}{y_2} S(1-z,\mu_s,\mu_f) f_{g/p} (y_1,\mu_f) f_{g/p} (y_2,\mu_f)  \nonumber \\
\label{sca}
&=& \tau H(M_S,\mu_s) \int^1_{\tau} \frac{dz}{z} S(1-z,\mu_s,\mu_f)  F(\tau/z,\mu_f),  
\end{eqnarray} 
where $\tau = M_S^2/s$, and $F(x,\mu)$ is a convolution of gluon PDFs, which can be written as 
\begin{equation} 
\label{lumi} 
F(x,\mu_f) =\int^1_{x} \frac{dy}{y} f_{g/p} (y,\mu_f) f_{g/p} (x/y,\mu_f).
\end{equation} 

\begin{figure}[t]
\begin{center} \vspace{-4cm}
\includegraphics[width=13cm]{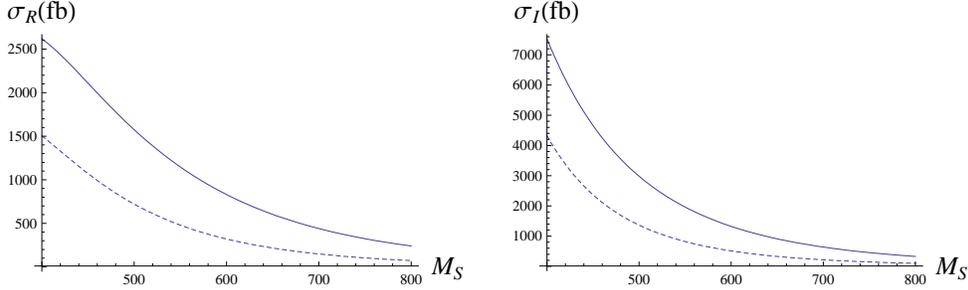} \vspace{-3.5cm}
\caption{\baselineskip 2.0ex 
The scattering cross sections employing Manohar-Wise Model at the LHC with $\sqrt{s}=14$ TeV. Here $\sigma_R$ and $\sigma_I$ indicate $\sigma(pp\to S_R X)$ and $\sigma(pp\to S_I X)$ respectively. In the both plots, the straight (dashed) lines denote the results at NLL (LO). 
}
\label{fig4}
\end{center}
\end{figure}

 Eq.~(\ref{sca}) is obtained by adopting the effective approach of (two-step) matching and running (see, e.g. \cite{Manohar:2003vb}). The hard function $H$ can be obtained from matching the (model dependent) physics above the hard scale hard $\mu_h \sim M_S$ onto effective operators in Soft Collinear Effective Theory (SCET) \cite{SCET1,SCETf}. The SCET fields represent the collinear gluons moving in opposite light-cone directions and the emission (in partonic threshold limit) of soft gluons into the final state. In that limit one also needs to consider the interactions of the heavy scalar octet with soft gluons much like in Heave Quark Effective Theory. After performing the first step matching  we then evolve down to $\mu_s\sim M_S (1-z)$ with the anomalous dimensions of the effective operators. $M_S(1-z)$ is the typical scale of the soft gluon radiation. Since the scale $\mu_s$ is still considered to be much larger than the non-perturbative scale, $\Lambda_{\rm QCD}$, we integrate out the radiated soft gluons (the second step matching) and then obtain the soft function $S$. Finally we evolve the soft function from $\mu_s$ to an arbitrary factorization scale $\mu_f$ with the DGLAP evolution kernel of the gluon PDFs.
Explicit expressions of the effective operators in SCET, their anomalous dimensions and exact definitions of the soft function (with one-loop results) will be given elsewhere \cite{idi}.

The above factorization theorem allows one to also perform resummation of large logarithms. This is done up to next-to-leading logarithms (NLL) using the technique of resummation which is directly performed in momenum space \cite{Resummation}.

Using Manohar-Wise Model \cite{Manohar:2006ga}, we give a prediction for the total scattering cross section. The Born approximation results are given in Ref.~\cite{Gresham:2007ri}. Using the Born results we computed the scattering cross sections for LHC at NLL. In Fig.~\ref{fig4}, we show the leading order (LO) and NLL results as function of  the octet scalar mass. For example, at $M_S = 500$ GeV, we obtain  
$\sigma_{\rm{NLL}} (pp \to S_R X) = 1.6$ pb and $\sigma_{\rm{NLL}} (pp \to S_I X) = 3.0$ pb, where $S_R$ and $S_I$ are the two neutral octet scalar fields obtained when decomposing the complex scalar octet field,  and we have chosen the input parameters, $\eta_U$ and  $\lambda_{4,5}$ as 1 following Ref.~\cite{Gresham:2007ri}. The NLL results are approximately twice larger than LO results, which are given as $\sigma_{\rm{LO}} (pp \to S_R X) = 0.7$ pb and $\sigma_{\rm{LO}} (pp \to S_I X) = 1.4$ pb at the same mass. The above results indicate that it is really possible to observe the single color-octet production at the LHC. It is also interesting to compare the results in Fig.~\ref{fig4} with the similar ones for the production of the SM Higgs given in \cite{Catani:2003zt}. The interesting point is that in the Higgs case, the impact of the NLL resummation is comparable to the next-to-leading order (NLO) one, however performing resummation is much easier than performing a complete NLO calculation as the latter requires considering two-loop Feynman diagrams. This is true for both the Higgs case and the scalar octet production. More discussion about that will be given in \cite{idi}.

 Other interesting possibilities for the production of color octet particles is the production of two such particles that might form a color singlet bound states (octetonia). Such scenario was discussed in \cite{Kim:2008bx}.
In Manohar-wise model the decay modes of the neutral color octet scalars depend crucially on the Yukawa couplings of that model. For certain region of the parameter space, the main decay mode would be the  $t\bar{t}$ channel \cite{Manohar:2006ga}.
 
In summary we have derived a general factorization theorem for the production of single color-octet state using SCET. The factorization theorem is universal: independent of new physics. The new physics dependence enters only in the the hard matching Wilson coefficient (the first step matching.) The factorized result for the total cross section looks similar to the SM Higgs production, however, and not surprisingly, the main difference will be the explicit form of the soft function which includes six soft Wilson lines compared to four in the Higgs case. By ressuming the large threshold logarithms, we estimated $\sigma (pp\to SX)$ up to NLL. Our results show that the higher order corrections are significant and are roughly twice larger than the LO cross section. Resummation up to next-to-next-leading logarithm (NNLL) is needed to see the suppression of the different scale dependences as well as to check the convergence of the perturbative expansion.

\end{document}